\documentclass[pre,twocolumn,showpacs]{revtex4-1}
\usepackage{graphicx}
\usepackage{bm}
\usepackage{amsmath}
\usepackage[percent]{overpic}
\begin{document}

\title{Fragmentation of brittle plates by localized impact}
\author{Rebeca C. Falc\~ao} 
\author{Fernando Parisio}
\email{corresponding author: parisio@df.ufpe.br} 
\address{Departamento de F\'{\i}sica, Universidade Federal de Pernambuco, 50670-901,
Recife, Pernambuco, Brazil}


\begin{abstract}
In this letter we address the fragmentation of thin, brittle layers due to the impact of high-velocity projectiles. Our approach is
a geometric statistical one, with lines and circles playing the role of cracks, randomly distributed over the surface.
The specific probabilities employed to place the fractures come from an analysis of how the energy input propagates 
and dissipates over the material. The cumulative mass distributions $F(m)$ we obtain are in excellent agreement 
with the experimental data produced by T. Kadono [Phys. Rev. Lett. {\bf 78}, 1444 (1997)]. Particularly, in the small mass regime we get $F(m)\sim m^{-\alpha}$, with $0.1<\alpha<0.3$ for a quite broad range of dissipation strengths and total number of fragments. In addition we obtain
the fractal dimension of the set of cracks and its correlation to the exponent $\alpha$ that account for the experimental results given by Kadono and Arakawa [Phys. Rev. E {\bf 65}, 035107(R) (2002)]. 
\end{abstract}
\maketitle

%

To a great extent, every day life is composed by a chain of non-equilibrium phenomena, which are hard to control or simulate, and whose concessions to theoretical modeling are quite limited. Apart from their relevance to basic research, many of these processes are of great technological importance, as for example, the breaking of a solid because of a sudden stress increase, to engineering and material science \cite{quiao,grady2, medvedovski,chen}. In spite of the progress in the field, the multifragmentation of solids is an emblematic illustration of the difficulties involved in a satisfactory physical description of far-from-equilibrium events.

Part of the trouble in dealing with fragmentation problems is the variety of possible scenarios. Measurable quantities, notably, the mass distribution of the fragments, are sensitive to the mechanical character of the material, e. g., brittle or ductile \cite{guin}, to its effective dimensionality (aspect ratios) \cite{bohr,meibom,linna, Ishii, gladden}, to its intrinsic geometry, e. g., a flat plate \cite{gomes, donangelo,donangelo2} or a spherical shell \cite{grady, her_eggs}, to the magnitude \cite{ching,myagkov} and spacio-temporal distribution of the energy input (uniform compression \cite{kadono1}, explosion \cite{grady, her_eggs}, projectile impact \cite{kadono1,kadono2} , etc). There is also a difficulty of distinct nature, namely, a considerable gap between geometric statistical models and what one could refer to as first principle fracture theories \cite{grady}. While the former involves the random positioning of lines on a surface (commonly straight lines on a flat surface) in the case of two-dimensional fragmentation, the latter intends to incorporate part of the physical processes that take place in the nucleation and propagation of cracks. 
\begin{figure}[ht!]
\includegraphics[width=3.6cm,angle=0]{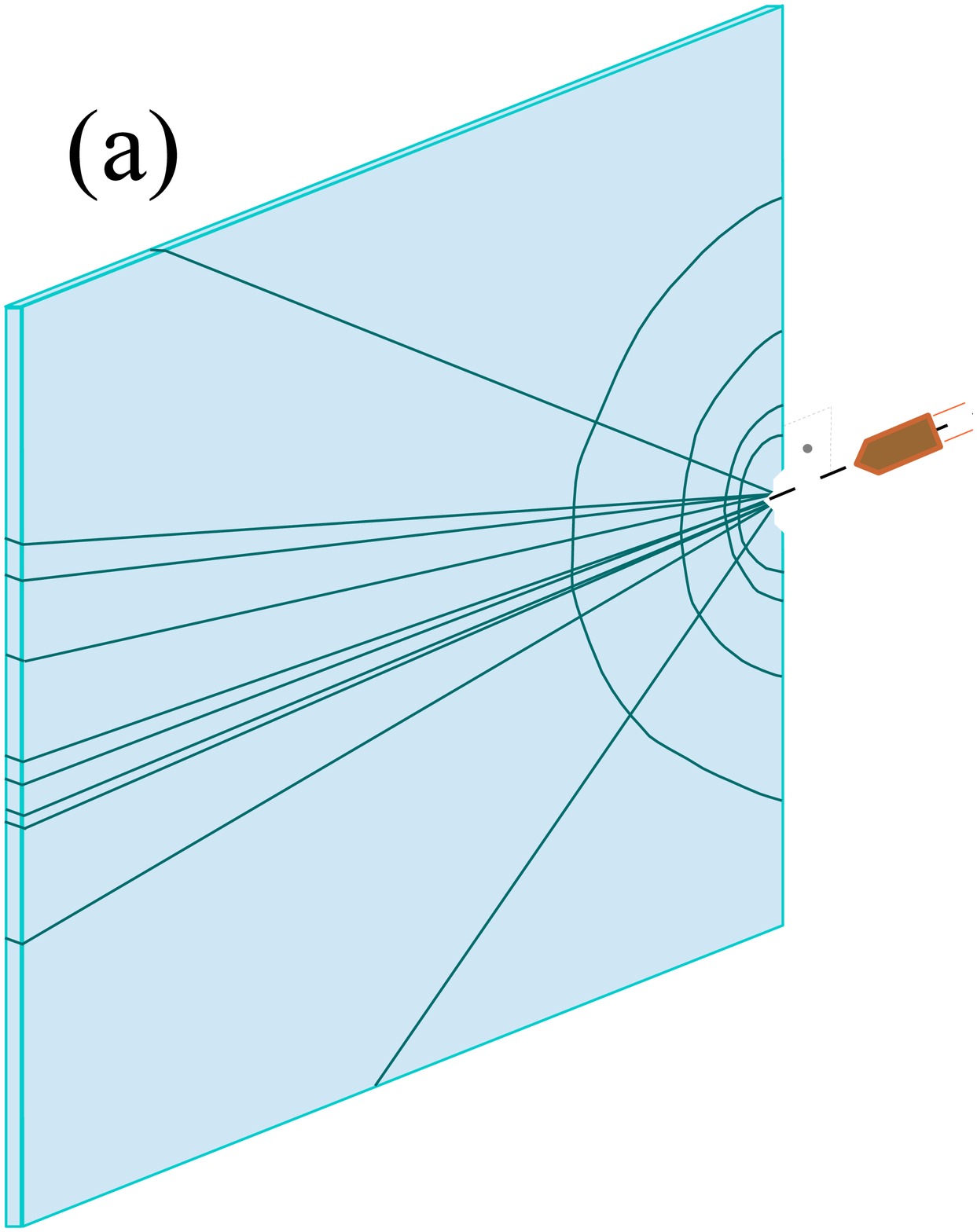}
\includegraphics[width=3.4cm,angle=0]{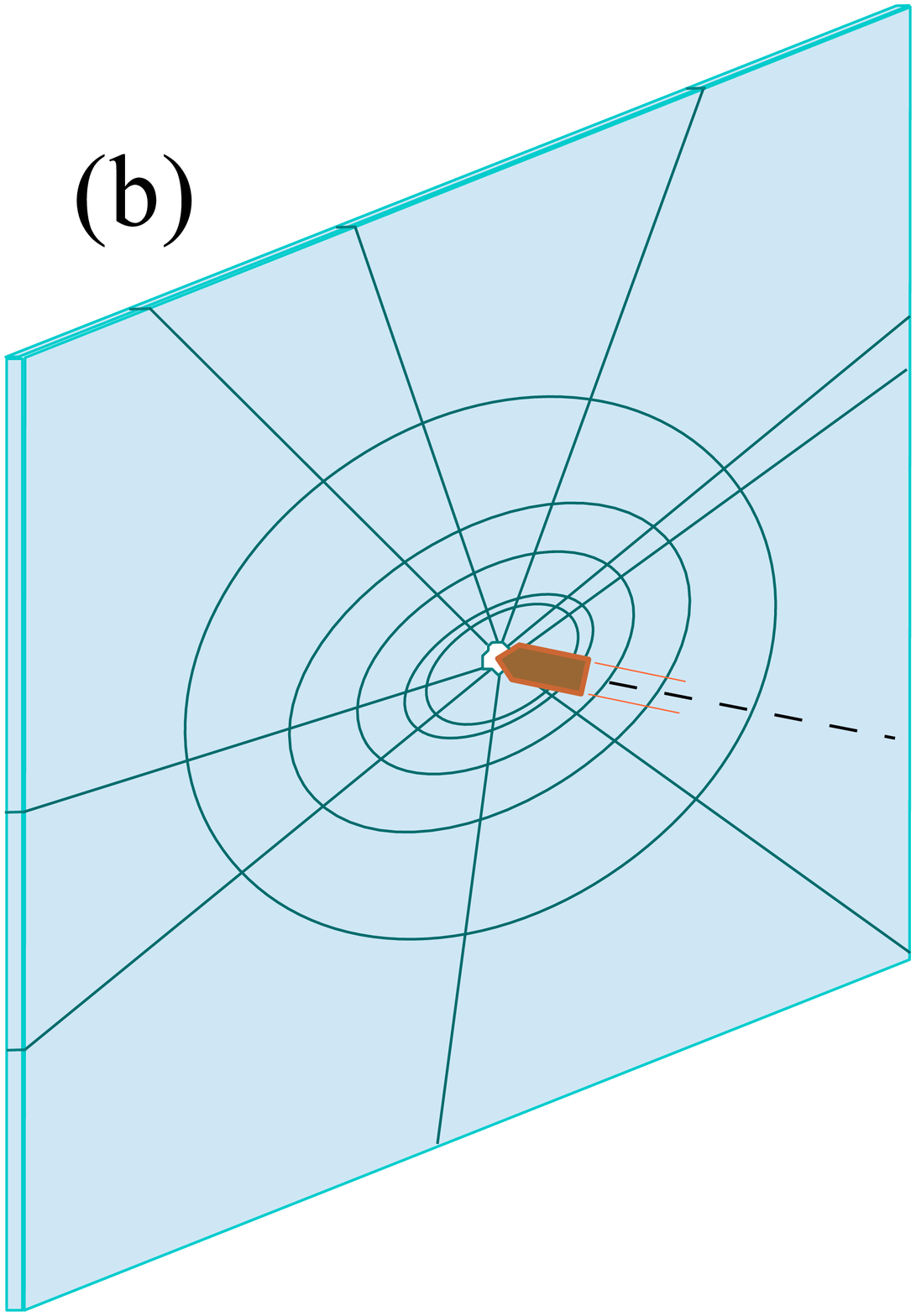}
\caption{(color online) Pictorial representation of the cracks formed on a brittle plate by localized lateral (a) and frontal (b) impacts. The projectiles and their trajectories are shown prior to the collision.}
\label{figure1}
\end{figure}

\begin{figure}
\includegraphics[width=6.5cm,angle=0]{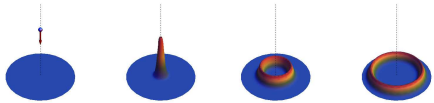}
\caption{(color online) Representation of the energy front as it propagates from the impact point (frontal incidence).}
\label{figure2}
\end{figure}

The present work concerns the particular, though important case of fragmentation of a flat brittle plate under a localized impact of a high-velocity projectile. We will address two specific situations, lateral impact [Fig. 1(a)], for which a variety of experimental data are available, and frontal impact perpendicular to the plate [Fig. 1(b)]. 
We propose an ensemble model based on geometric statistics which considers the observed disposition of the cracks and, importantly, energy transport and dissipation. We argue that the most relevant difference between spatially uniform versus localized energy inputs is that in the first case the energy is more or less evenly delivered to each part of the surface (no energy transport needed over macroscopic distances), while in the second case the energy is delivered at a small region and then propagates over the material, see Fig. \ref{figure2}. The transport of mechanical energy over macroscopic distances certainly implies losses, that will be shown to play a crucial role. Before describing our model we discuss some key experiments.

By far, the most commonly reported quantity in experiments and simulations is the relative number of fragments with mass larger than $m$, given by
$F(m)=\int_m^Mp(m'){\rm d}m'$,
where $M$ is the total mass of the plate and $p(m)$ is the probability density to get a fragment with mass between $m$ and $m+$d$m$. In fragmentation processes in general, and specifically in the situations we are interested in, it is common to have
\begin{equation}
F(m)\sim m^{-\alpha}\;,
\end{equation}
for small fragments, although the distribution as a whole is often better described by a power law with an exponential cutoff $F(m)\sim m^{-\alpha}\exp(-m/m_0)$. More intricate behaviors are also observed as, for example, composite power laws \cite{gomes,donangelo,donangelo2,meibom}. 

Regarding the magnitude of the energy intake, the structures that are usually formed, in the case of orthogonal incidence, were described in \cite{astrom}: ``The radial cracks were fairly straight and directed outwards from the point of impact, while the tangential cracks formed a more or less circularly symmetric crack with the impact point as the center of the circle. At still lower impact velocities, only radial cracks were formed.'' Recently, N. Vandenberghe et al have studied this same class of problems in a range of velocities where only radial cracks appear \cite{van}. 

In the case of lateral incidence, although the cracks may display the same kind of geometry, the patterns may also be more elaborate as is the case of Fig. 1(b) of \cite{kadono1}. Still, the author observes that large cracks are either radial or perpendicular to the radial ones. The set of snapshots in Fig. 2 of \cite{kadono2} presents a clearer geometry. In \cite{kadono1} Kadono presents an experimental investigation on plaster and glass disks laterally impacted. The nylon projectiles are accelerated by a light-gas gun to velocities around $4$ Km/s, and the focus is on the mass distribution of the produced fragments. For aspect ratios $\gamma=$diameter/thickness$ >60$ (2D regime) the author obtained $0.1<\alpha<0.3$ in all realizations. This narrow interval is quite characteristic of localized impact, as we will see. In the same reference uniform compression essays (``sandwich'' experiments) were also carried out and the obtained exponents satisfy $0.5<\alpha<0.7$. In a subsequent article, Kadono and Arakawa \cite{kadono2} executed lateral impact essays with projectile velocities up to 67 m/s, impinging on transparent plates. By using a stroboscopic camera the authors were able to estimate the fractal dimension of the cracks on the target as a function of time. Although these works have been often cited in the literature, a statistical model that correctly accounts for the fragment mass distribution and corresponding exponents, its connection to the fractal dimension of the set of fractures, and the robustness of the exponents against variations of the energy input, is missing. To provide such a model is the main goal of the present letter.

Our procedure is to consider ensembles of $10^3$ samples, each fragmented in a fixed number of pieces $N$. Although we used different values of $N$ (100 to 900), most of the experimental results involve a few hundred of fragments, typically less than 300. 
In fact, the obtained exponents were not very sensitive to the value of $N$ in the investigated range, which confirms the experimental observation that the interval of exponents remains approximately unchanged when the impact velocity varies by a factor as large as 50 \cite{kadono1}. 
%

%
%
\begin{figure}
\includegraphics[width=2.8cm,height=5cm,angle=0]{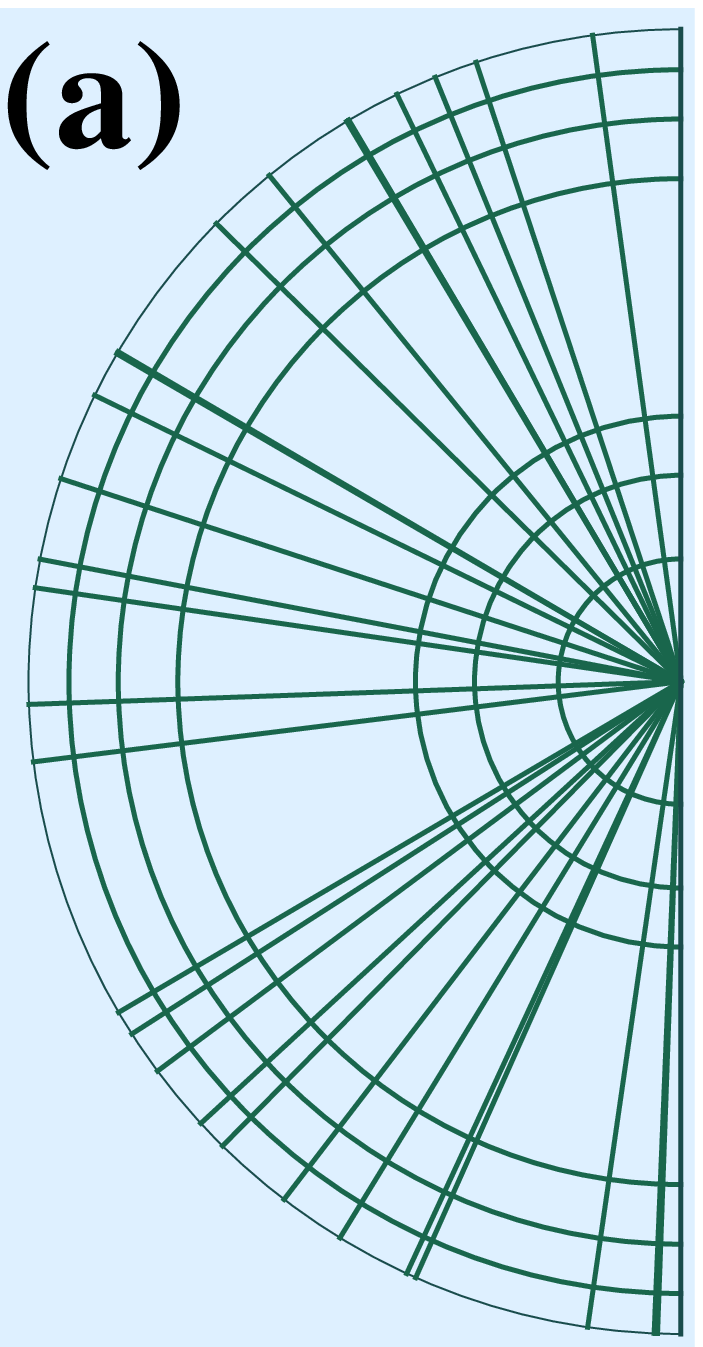}
\includegraphics[width=2.8cm,height=5cm,angle=0]{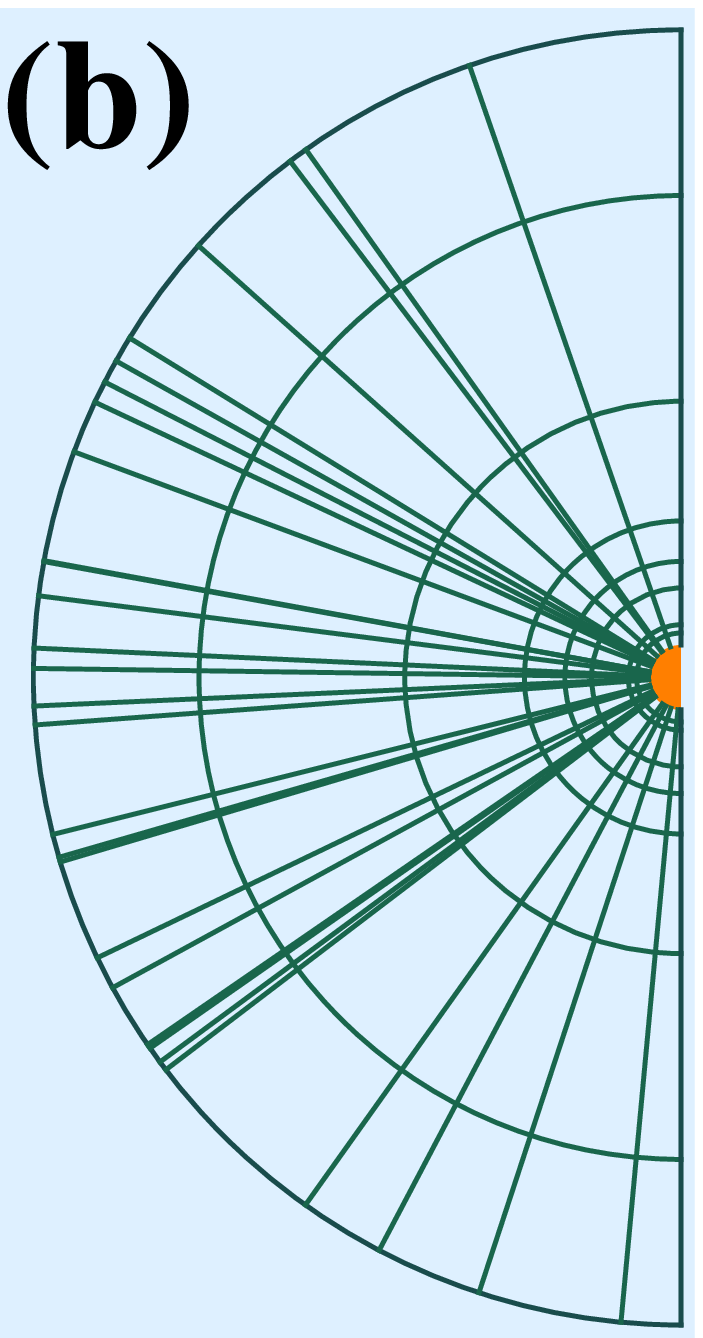}
\caption{(color online) Panel (a) shows a typical pattern generated by a uniform distribution of circular cracks and a isotropic distribution of radial fractures. In (b) the pattern generated by our model is depicted. Note that the circular cracks are more concentrated around the impact point. The number of radial and circular cracks are the same in both panels.}
\label{figure3}
\end{figure}

In order to emphasize the role of energy transport and dissipation in localized impact fragmentation, we begin with an oversimplified model that only accounts for the overall geometric aspect of the cracks. Let us consider an ensemble of homogeneous identical half-disks of radius $R$ and assume that the total number of fragments is fixed and given by $N=(N_r+1)(N_c+1)$, where the cracks are represented by $N_r$ radial segments (emerging from the impact point) and $N_c$ concentric semi-circles (centered at the impact point). Let $p(\theta)$ and $p(r)$ be the probability densities to find a radial crack with angular position between $\theta$ and $\theta +$d$\theta$ and to find a circular crack between $r$ and $r+$d$r$ apart from the central point, respectively. We take an ensemble where the positioning of both kinds of crack obeys the simplest distributions, that is, $p(\theta)=1/\pi$ and $p(r)=1/R$. See figure \ref{figure3}(a) for a typical pattern with $N_r=29$ and $N_c=7$. Since, for low/intermediate energies only radial fractures are produced, it is more usual to have $N_r>N_c$. Note the visual aspect of the pattern, with cracks more or less evenly distributed. We found that the cumulative function $F(m)$ does not follow a power law, being completely incompatible with the available experimental data, making it clear that some essential ingredients are missing.
%
%
The model we propose here is based the following assumptions:

{\bf (i)} We consider an ensemble of homogeneous identical half-disks (disks) of radius $R$, laterally (perpendicularly) impacted by projectiles of same size and mass. The cracks are represented by $N_r$ radial segments and $N_c$ concentric semi-circles (circles) centered at the impact point. The total number of fragments is given by $N=(N_r+1)(N_c+1)$ [$N=N_r(N_c+1)$].

{\bf (ii)} We assume that a small semi-circular (circular) region with radius $r_0$ is shattered by the impact. The fragment statistics is considered for $r \in [r_0,R]$ \cite{comm}.

{\bf (iii.a)} In the case of lateral impact, the probability that, over the ensemble, a radial fracture is found between $\theta$ and $\theta+$d$\theta$ is $p(\theta)$d$\theta$, with $p(\theta) \propto\cos \theta$, where $\theta$ is the angle between the radial crack and the incidence direction of the projectile. We verified that the particular form of $p(\theta)$ is of little importance, provided that the failures concentrate near the original projectile's direction.

{\bf (iii.b)} For frontal impacts, isotropy is a natural assumption and the probability that a radial fracture is found between $\theta$ and $\theta+$d$\theta$ is $p(\theta)$d$\theta$, with $p(\theta)=1/2\pi$. 

{\bf (iv)} Part of the energy transferred by the projectile is almost instantly employed to form the central hole and then the radial fractures. The remaining energy, that we denote by $E_0$, propagates through the material forming the circular (or tangential) cracks.

{\bf (v)} The distribution of circular fractures in the ensemble, $p(r)$, satisfies to first order $p(r)\propto \rho(r)$, where $\rho$ is the energy density a distance $r$ away from the impact point, carried by the wave front generated by the impulsive energy intake. When losses are neglected we have $\rho(r)=E_0/2\pi r$. However, as the disturbance caused by the impact propagates on the plate, energy is absorbed due to heat conversion, creation of phonons, etc. Given the macroscopic homogeneity of the material, we suppose that the smallest area relevant to our problem is amenable to coarse-graining arguments, and thus, the {\it fraction} of energy absorbed per unit area is constant. We assume that this is the prevailing dissipation mechanism for brittle materials. The energy necessary to produce the cracks themselves is already present in the material in potential form and is released by the energy pulse \cite{grady2}. We define a restitution parameter $q$ such that $E(R)=E_0$ for $q=1$ and $E(R)=0$ for $q=0$, while $E(r_0)=E_0$ for arbitrary $q$. This leads to
$\frac{E(r)}{E_0}=1-\frac{1-q}{1-\zeta^2}\left[ 1-\left( \frac{r_0}{r}\right)^2\right]$,
where $\zeta=r_0/R$. The energy density over the wave front is
$\rho=E(r)/2\pi r$, leading to the final, normalized result for the probability distribution of tangent cracks in the ensemble: 
\begin{equation}
p(r)=\frac{1-\zeta^2-(1-q)\left[ 1-\left( \frac{r_0}{r}\right)^2\right]}{r\left[ \frac{1-q}{2}+\left(\frac{1-q}{1-\zeta^2}+1\right)\ln\zeta\right]}\;.
\label{dristr_r}
\end{equation}
In Fig. \ref{figure3} (b) we show the pattern formed when our model is applied for $N_r=29$, $N_c=7$, and $q=0.3$. Note that most of the circular cracks are concentrated around the impact point when dissipation is taken into account and the radial fractures propagate with higher probability in the horizontal direction. If the projectile is sizable, as in \cite{kadono2}, the fractures emanate with higher probability from the edges of the projectile. 

%
\begin{figure}
\includegraphics[width=3.9cm,angle=0]{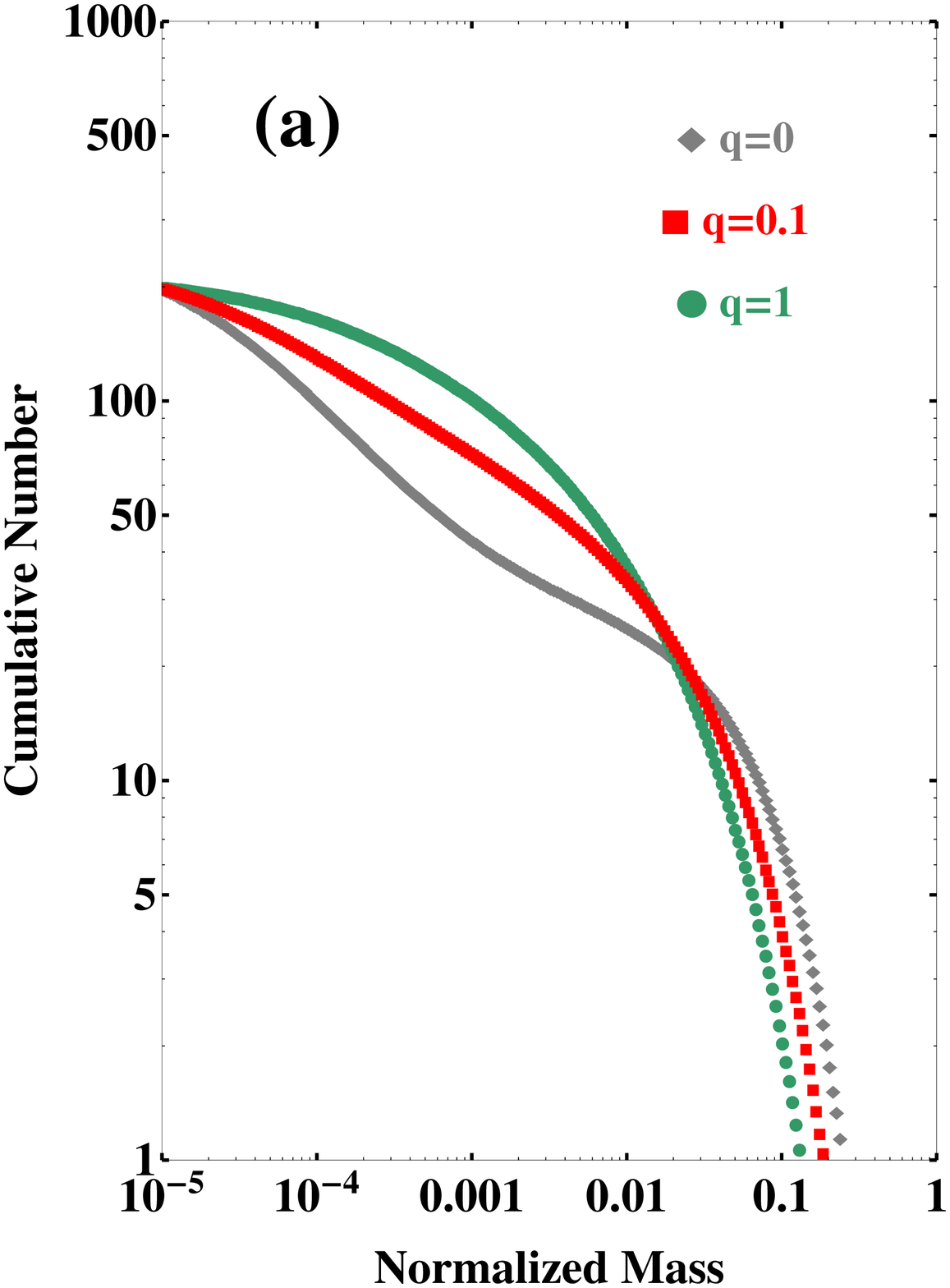}
\includegraphics[width=4.1cm,angle=0]{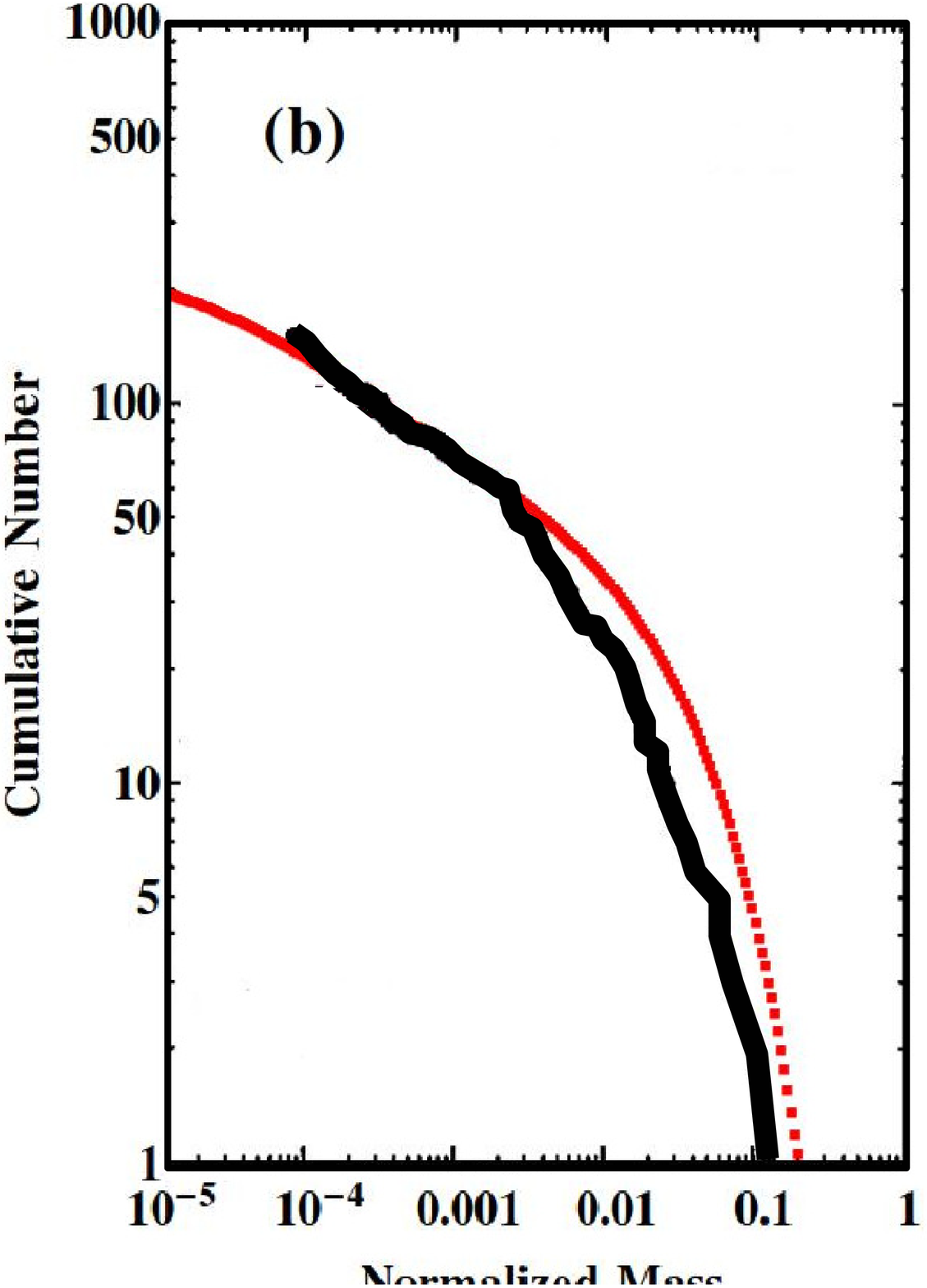}
\caption{(color online) (a) Cumulative number [$N\times F(m)$] for $N=200$ and $q=0$, $q=0.1$, and $q=1$. (b) Results from figure 2(c) of reference \cite{kadono1} and our curve for $q=0.1$. }
\label{figure4}
\end{figure}
The presented model produces cumulative mass distributions whose form is strongly supported by the experimental results in the whole range of masses. In the small fragment regime, power-law behaviors appear when dissipation is substantial. In Fig. \ref{figure4} (a) we plot $F(m)$ for a total of $200$ fragments (with $N_r=19$ and $N_c=9$) and $q=0$, $q= 0.1$, and $q=1.0$. The fragment number was chosen to fit the experimental plot presented in figure 2(c) of reference \cite{kadono1}. In Fig. \ref{figure4} (b) we superimpose our $F(m)$ with $q=0.1$ as it stands in Fig. \ref{figure4} (a) and the result presented in \cite{kadono1}. The scales are exactly the same. Note, in addition, that while the curve for $q=0$ (total dissipation) has a concavity that is not observed in the experimental plots, the curve for the conservative case ($q=1.0$) does not present a power-law behavior, showing that dissipation is a fundamental ingredient to reproduce the results obtained in the laboratories.
\begin{figure}
\includegraphics[width=4.2cm,angle=0]{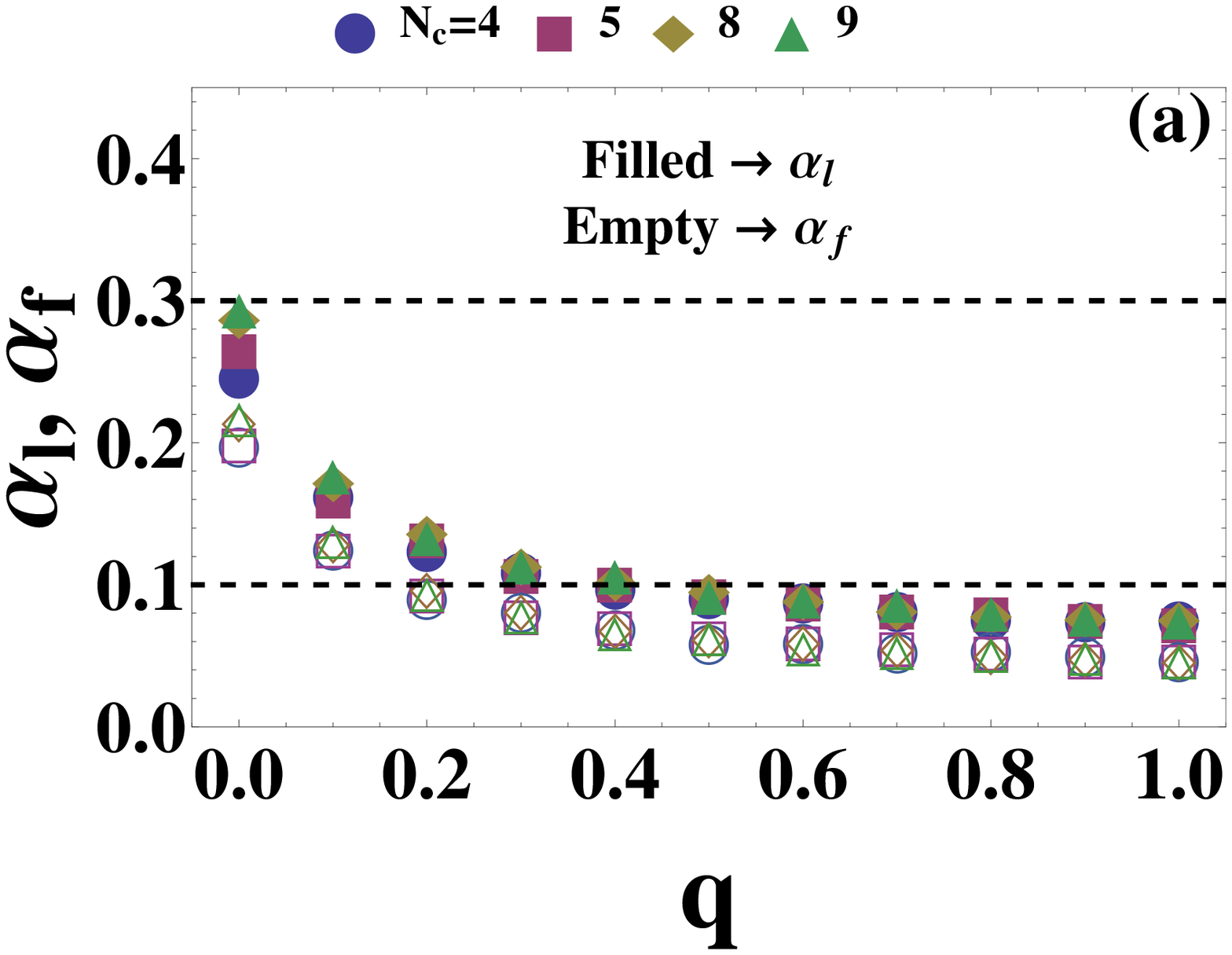}
\includegraphics[width=4.2cm,angle=0]{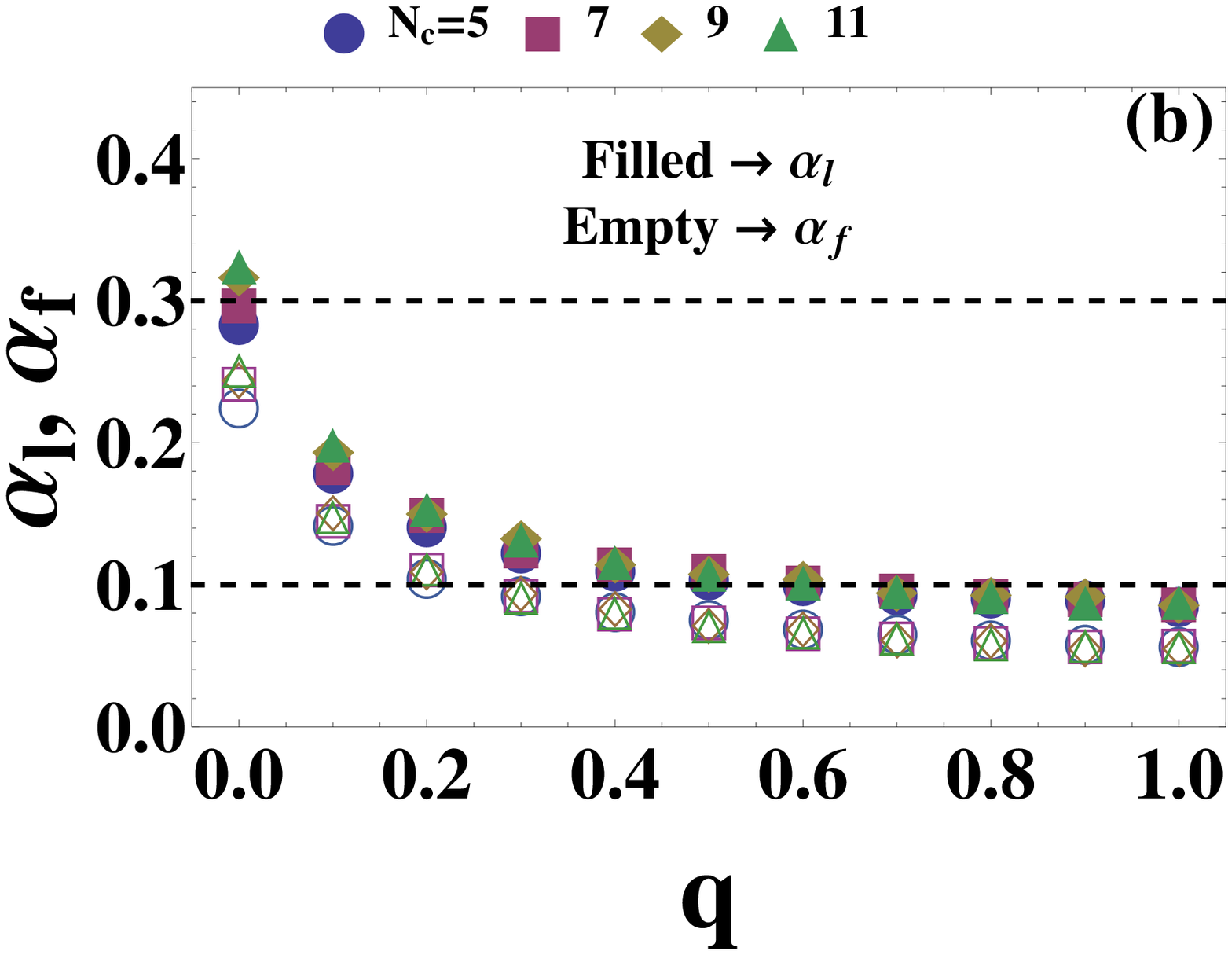}
\includegraphics[width=4.2cm,angle=0]{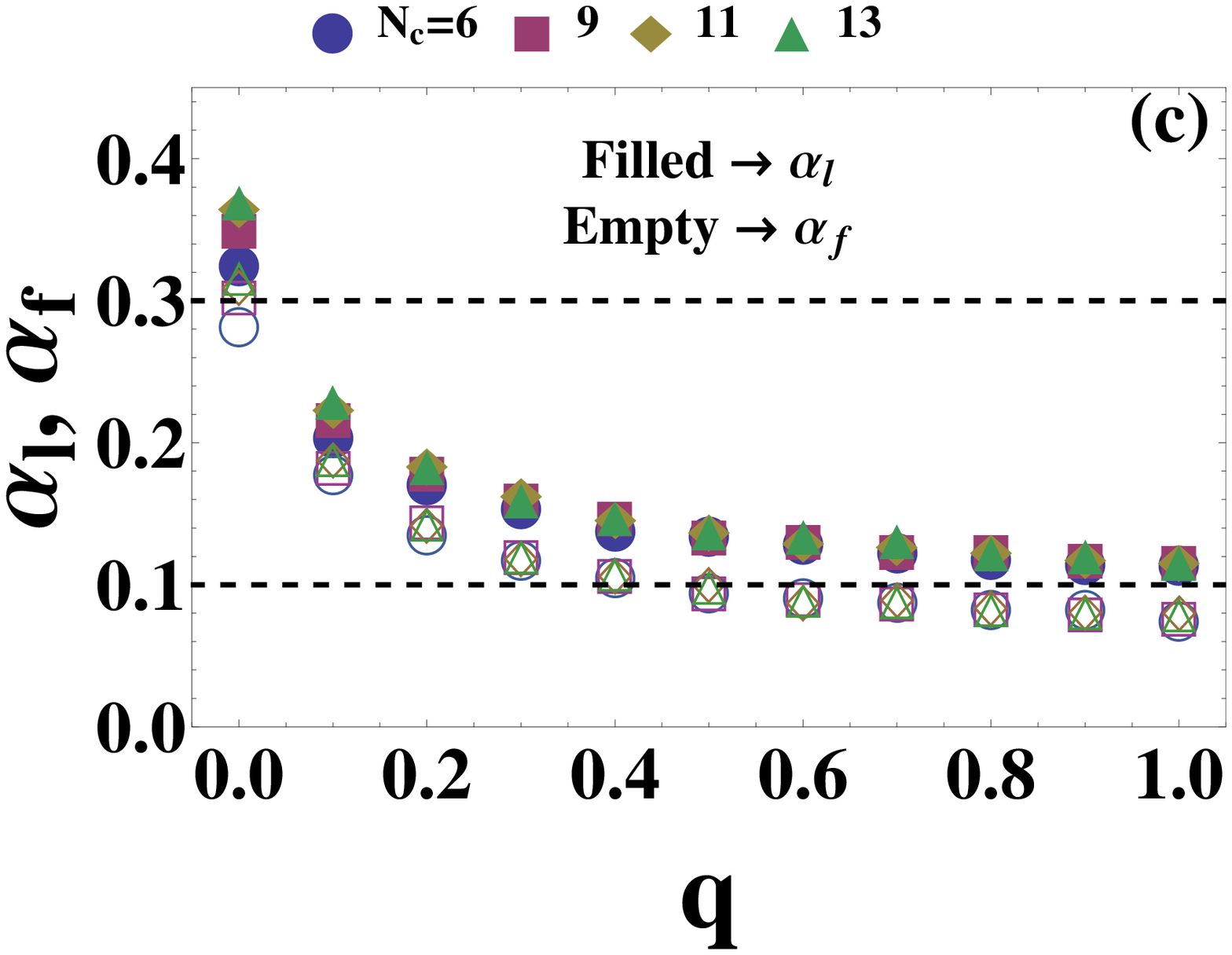}
\includegraphics[width=4.2cm,angle=0]{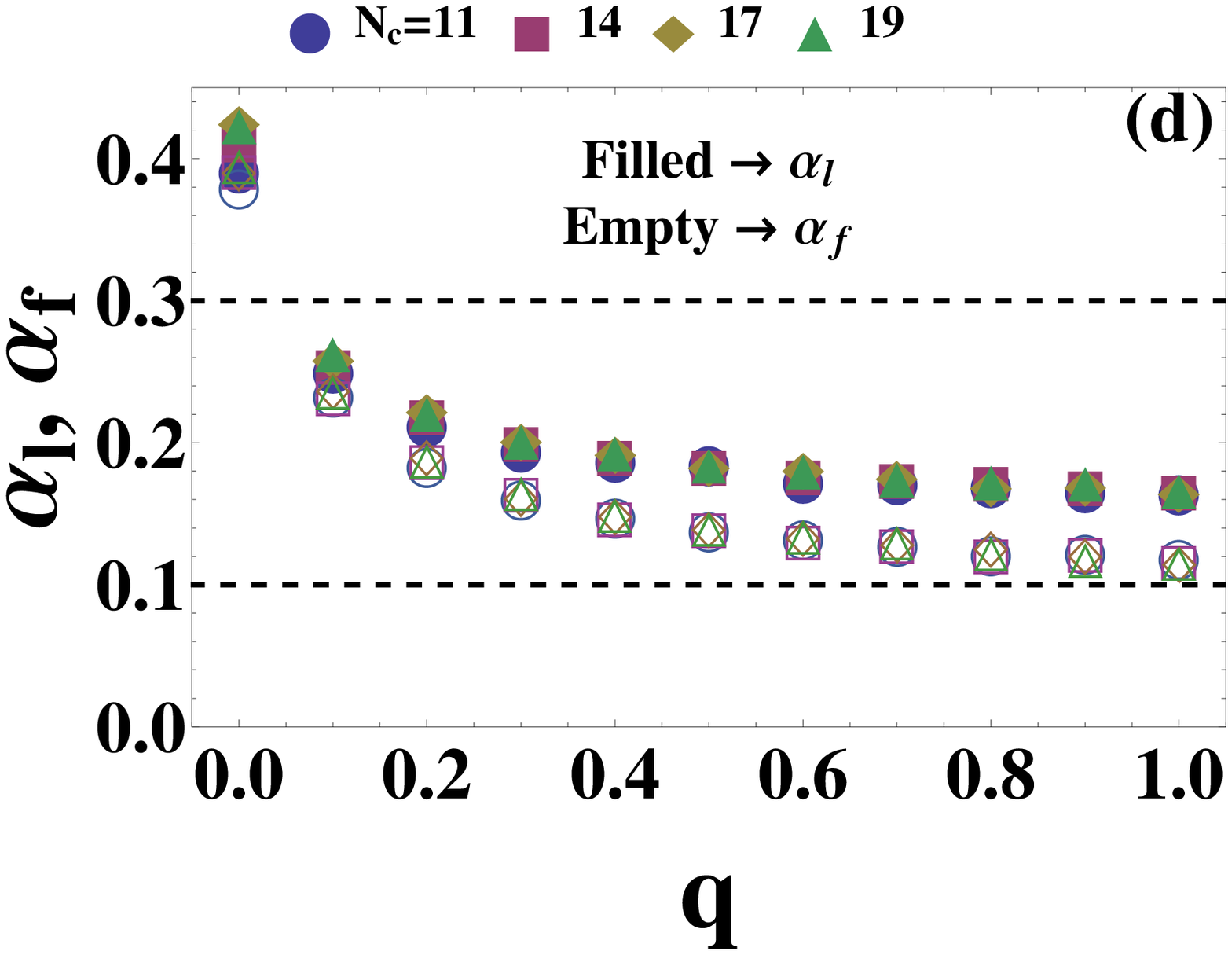}
\caption{(color online) Exponent $\alpha_l$ and $\alpha_f$ versus the restitution parameter $q$ for $N=180$ (a), $N=240$ (b), $N=420$ (c), and $N=900$ (d) fragments. Filled symbols refer to lateral impact ($\alpha_l$) and empty symbols refer to frontal impact ($\alpha_f$). The two dashed horizontal lines give the interval of experimental values obtained for $\alpha_l$ in \cite{kadono1}.}
\label{figure5}
\end{figure}
Although the visual match may be convincing, in order to check the overall predictions of the model we must evaluate the exponents it produces and compare them to those obtained in experiments for a wide range of values of $q$, $N$ and of the ratio $N_r/N_c$. These data are gathered in Fig. \ref{figure5} where we show the exponents $\alpha_l$ (lateral impact) and $\alpha_f$ (frontal impact) versus the restitution parameter $q$ for $N=180$ (a), $N=240$ (b), $N=420$ (c), and $N=900$ (d), for a variety of values of $N_r$ and $N_c$ (keeping $N$ unchanged). Filled symbols refer to lateral impact and empty symbols refer to frontal impact. The two dashed horizontal lines give the interval of experimental values obtained for $\alpha_l$ in \cite{kadono1}. The first thing that stands out is that the interval $[0.1, 0.3]$ contains almost all points for lateral impact in all cases. We stress that all the points referring to higher values of $q$, typically $q>0.5$, are not associated to genuine power laws (valid for at least one decade) and are plotted for completeness. In this case, one can only speak of average {\it slopes} of the distributions in the small mass domain [see Fig. \ref{figure4} (a)]. In references \cite{kadono1,kadono2} the typical number of fragments is compatible to those used in figures 5 (a) and 5 (b), from where we see that dissipation plays a key role in our model, and, arguably, in the laboratory essays. The lowest experimental exponents are reproduced for $q \approx 0.5$ ($N=240$), while the value $\alpha_l=0.3$ is reached for strongly dissipative scenarios, with $q\approx 0$. For larger numbers of fragments all the exponents are shifted upwards, but in the majority of the cases they remain between the experimental limits [see figs 5 (c) and 5 (d)]. This supports the observation that the interval of exponents is almost insensitive to the variation of the projectile velocity over more than one order of magnitude. Note that in all cases the value of the ratio $N_r/N_c$ plays a quite secondary role. As expected, due to isotropy, the situation of frontal impact leads to slightly smaller exponents ($\alpha_l>\alpha_f$). 
\begin{figure}[!ht] 
\begin{overpic}[width=0.45\textwidth]{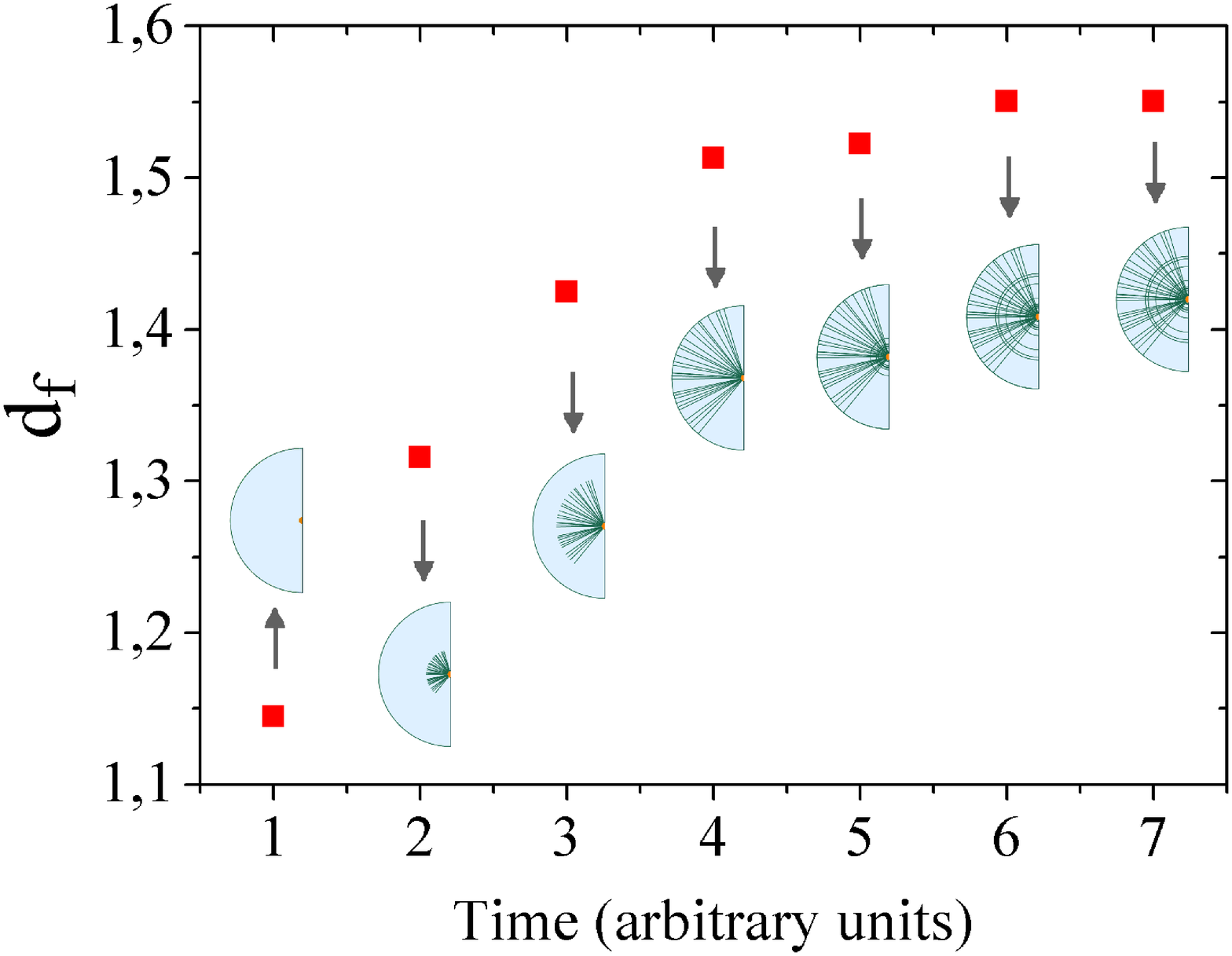}
\put(63,15){\includegraphics[width=35\unitlength]{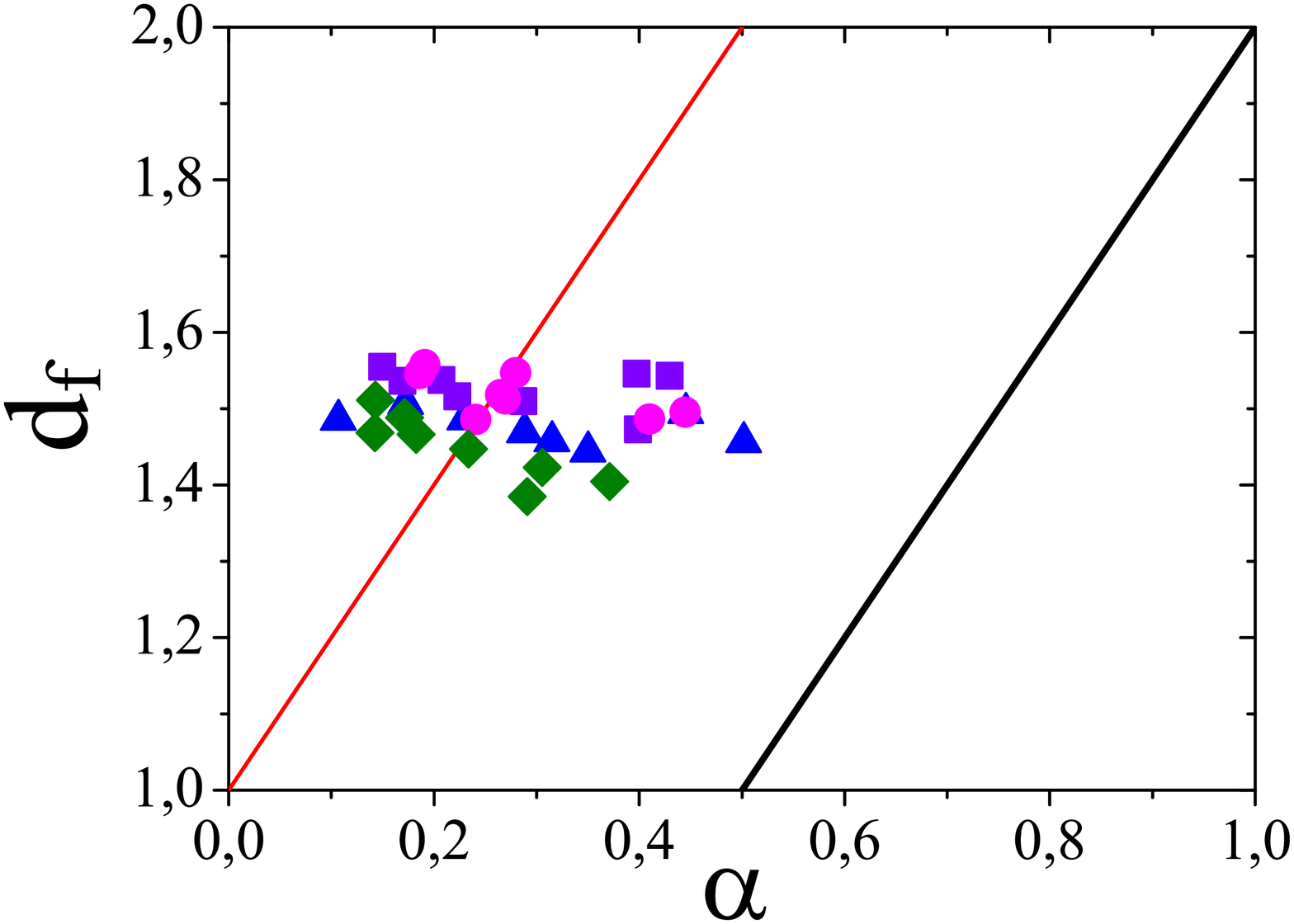}} 
\end{overpic}
\caption{(color online) Time development of the box counting dimension. The inset shows the box count dimension versus the exponent $\alpha$.}
\label{figure6}
\end{figure}

We finally address the fractal dimension of the set of cracks on the target while the pattern develops. This problem has been experimentally investigated in \cite{kadono2}. In this reference the authors find a steep increase in the fractal dimension ($d_f$), as the first snapshots are considered, and then, in a second stage, a much slow increase with a saturation around $d_f=1.55$. By noticing that in the snapshots shown in Fig. 2 of \cite{kadono2} the radial fractures are formed first (panels 1 to 3) and only later the tangential cracks develop (panels 4 to 6), both at a constant velocity of the order of the speed of sound in Pyrex glass ($\sim 5.6$ km/s) \cite{comment3}, we start from the plate with the damaged region only followed by six panels. First we let radial cracks grow at a constant rate being followed by the later circular cracks. The result of the box counting dimension for each step is presented in Fig. \ref{figure6}. The agreement with the experimental plot, Fig 3 (a) of \cite{kadono2} is quite accurate, both qualitatively and quantitatively. In addition, we remark that the last three snapshots (7 to 9) in Fig. 2 of \cite{kadono2} seem to show the effect of a secondary impact (due to the reaction by the support). The inset, to be compared with Fig. 5 of \cite{kadono2}, shows the final box counting dimension versus the exponent $\alpha$ for $q \in [0,0.5]$ and $N=180,240,420$ fragments for various ratios $N_r/N_c$. The results are compatible to those obtained in \cite{kadono2} while we were not able to obtain the higher exponents. Perhaps our model is too simple to produce these exponents. Another tenable possibility is that the values of $\alpha$ might have been increased by both, the secondary impact and the sizable cylindrical projectile (in this case the radial cracks tend to propagate diagonally from the two edges of the projectile). The two straight lines, $d_f=2\alpha$ (left) and $d_f=2\alpha+1$ (right), refer to simplified models proposed by the authors of \cite{kadono2}, and are shown to guide the eye.

%

In this letter we presented a macroscopic model for impact fragmentation that led us to obtain the probability distributions of fractures analytically. We found that the experimental observations regarding the cumulative distributions of fragment masses, the power-law behavior for small masses, the interval of exponents $\alpha$, its relative insensitivity to changes in the energy input, as well as the quasi-fractal properties of the set of cracks can be understood provided that energy transport and dissipation are considered together with the spider web-like geometry of the fractures. An immediate perspective is to apply the ideas presented here to uniform lateral impact essays, like those reported in \cite{donangelo,donangelo2}. 

\begin{acknowledgements}
The authors thank the comments and suggestions by M. A. F. Gomes (DF-UFPE) and K. Coutinho (IF-USP). Financial support from Conselho Nacional de Desenvolvimento Cient\'{\i}fico e Tecnol\'ogico (CNPq), Coordena\c{c}\~ao de Aperfei\c{c}oamento de Pessoal de N\'{\i}vel Superior (CAPES), and Funda\c{c}\~ao de Amparo \`a Ci\^encia e Tecnologia do Estado de Pernambuco (FACEPE) (Grant No. APQ-1415-1.05/10) is acknowledged.
\end{acknowledgements}


\begin{thebibliography}{99}
\bibitem{comm} In this letter we set the ratio $\zeta=r_0/R=0.05$ in all presented results. We did test our model for different values, for instance $\zeta=0.1$, obtaining almost identical numerical results.
\bibitem{comment3} The side length of the square Pyrex target is 100 mm and the time interval between successive snapshots is 10 $\mu$s.
\end{thebibliography}
\end{document}